%% file: paper1.tex
\documentclass[twocolumn,showpacs,preprintnumbers,amsmath,amssymb,nofootinbib,superscriptaddress]{revtex4}

\usepackage{graphicx,amssymb,amsmath,multirow,amsfonts,amsthm,latexsym,graphics}% Include figure files
\usepackage{dcolumn}% Align table columns on decimal point
\usepackage{bm,color}% bold math
\usepackage{hyperref}

\input{inputs/abbr-journals}

\newcommand{\Omegam}{\Omega_{\rm m}}

\newcommand{\Omegade}{\Omega_{\rm de}}
\newcommand{\Omegab}{\Omega_{\rm b}}
\newcommand{\dd}{{\rm d}}
\newcommand{\ontop}[2]{
  \renewcommand{\arraystretch}{0.2}
  \begin{array}{c}
  #1 \\ #2
  \end{array}
  \renewcommand{\arraystretch}{1.0}
}
\newcommand{\lsim}{\ontop{<}{\sim}}

\newcommand{\transp}{{\rm T}}
\newcommand{\indicator}{{\mathbf 1}}

\begin{document}
%%%%%%%%%%%%%%%%%%%%%%%%%%%%%%%%%%%%%%%%%%%%%%%%%%%%%%%%%%%%

\preprint{}

%%%%%%%%%%%%%%%%%%%%%%%%%%%%%%%%%%%%%%%%%%%%%%%%%%%%%%%%%%%%
\title{Estimation of cosmological parameters using adaptive
  importance sampling}
%%%%%%%%%%%%%%%%%%%%%%%%%%%%%%%%%%%%%%%%%%%%%%%%%%%%%%%%%%%%

\author{Darren Wraith}
\affiliation{CEREMADE, Universit\'e Paris Dauphine, 75775 Paris cedex 16, France}
\affiliation{Institut d'Astrophysique de Paris, CNRS UMR 7095 \&
  UPMC, 98 bis, boulevard Arago, 75014 Paris, France}

\author{Martin Kilbinger}
\affiliation{Institut d'Astrophysique de Paris, CNRS UMR 7095 \&
  UPMC, 98 bis, boulevard Arago, 75014 Paris, France}
\author{Karim Benabed}
\affiliation{Institut d'Astrophysique de Paris, CNRS UMR 7095 \&
  UPMC, 98 bis, boulevard Arago, 75014 Paris, France}

\author{Olivier Capp\'e}
\affiliation{LTCI, TELECOM ParisTech and CNRS, 46, rue Barrault,
  75013 Paris, France}

\author{Jean-Fran\c{c}ois Cardoso}
\affiliation{LTCI, TELECOM ParisTech and CNRS, 46, rue Barrault,
  75013 Paris, France}
\affiliation{Institut d'Astrophysique de Paris, CNRS UMR 7095 \&
  UPMC, 98 bis, boulevard Arago, 75014 Paris, France}

\author{Gersende Fort}
\affiliation{LTCI, TELECOM ParisTech and CNRS, 46, rue Barrault,
  75013 Paris, France}

\author{Simon Prunet}
\affiliation{Institut d'Astrophysique de Paris, CNRS UMR 7095 \&
  UPMC, 98 bis, boulevard Arago, 75014 Paris, France}

\author{Christian P.~Robert}
\affiliation{CEREMADE, Universit\'e Paris Dauphine, 75775 Paris cedex 16, France}

\date{\today}% It is always \today, today,
             %  but any date may be explicitly specified

\begin{abstract}

We present a
Bayesian sampling algorithm called adaptive importance sampling or
Population Monte Carlo (PMC), whose computational workload is easily
parallelizable and thus has the potential to considerably reduce the
wall-clock time required for sampling, along with providing other
benefits. To assess the performance of the approach for cosmological
problems, we use simulated and actual data consisting of CMB
anisotropies, supernovae of type Ia, and weak cosmological lensing,
and provide a comparison of results to those obtained using
state-of-the-art Markov Chain Monte Carlo (MCMC).  For both types of
data sets, we find comparable
parameter estimates for PMC and MCMC, with the advantage of a
significantly lower computational time for PMC.  In the case of WMAP5
data, for example, the wall-clock time reduces from several days for
MCMC to a few hours using PMC on a cluster of processors.  Other
benefits of the PMC approach, along with potential difficulties in
using the approach, are analysed and discussed.

\end{abstract}

% PACS, the Physics and Astronomy
                             % Classification Scheme.
\maketitle

%%%%%%%%%%%%%%%%%%%%%%%%%%%%%%%%%%%%%%%%%%%%%%%%%%%%%%%%%%%%
\section{Introduction}
\label{sec:intro}
%%%%%%%%%%%%%%%%%%%%%%%%%%%%%%%%%%%%%%%%%%%%%%%%%%%%%%%%%%%%

In recent years we have seen spectacular advances in observational
cosmology, with the availability of more and more high quality data
allowing for the testing of models with higher complexity.  Some of
these tests have been made possible thanks to the use of Bayesian
sampling techniques, and in particular Markov Chain Monte Carlo
(MCMC) -- an (iterative) algorithm that produces a Markov chain whose
distribution converges to the target posterior $\pi$.  After a
``burn-in'' period, samples from such a chain can be regarded as
samples approximately from $\pi$. Proposed values for the chain or the
updating scheme of MCMC can be designed to ensure that moves towards
regions of higher mass under $\pi$ are favored, and regions with null
probability (under $\pi$) are never visited. This way, most of the
computational effort can be spent in the region of importance to the
posterior distribution, and an MCMC approach is usually much more
efficient than traditional grid sampling of the model parameter space.

The MCMC technique is now well known in cosmology, and in particular
in its most simple form, the Metropolis-Hastings algorithm, thanks to
the user-friendly and freely available package
COSMOMC~\citep{cosmomc}.  Other forms of the MCMC algorithm, like
Gibbs-sampling and Hybrid Monte Carlo (better known in cosmology as
Hamiltonian sampling), have also been proposed and have found some
interesting usage in the estimation of the posterior distribution for
the Cosmic Microwave Background anisotropy power spectrum at low
resolution (see \citep{Rudjord:2008p4864} and references therein, also
\citep{Taylor:2007p3316} and \citep{WMAP5-Dunkley08}).

For all its advantages over grid sampling, the MCMC approach also
suffers from problems. One difficulty is to assess the correct
convergence of the chain. Another lies in the presence of correlations
within the chain which can greatly reduce the efficiency of the
sample~\citep{robert:casella:2004}. A third issue which is
particularly relevant for the usage of MCMC in cosmology is the
computational time involved. Indeed, whatever the sampling technique,
we often need to compute at least one estimate of the posterior for
each sampled point. This computation can be slow in cosmology. With
the current processing speed of computers, a point of the posterior
of, for example, the WMAP5 data set, using CAMB
\citep{Lewis:1999bs}\footnote{http://camb.info} and the public WMAP5
likelihood code
\citep{WMAP5-Dunkley08}\footnote{http://lambda.gsfc.nasa.gov}, both
with their default precision settings, is computed at the order of several
seconds, and can be much slower when exploring non-flat models. Of
course, as stated above, for most problems MCMC will require orders of
magnitude less samples than a grid for a given target precision, thus
providing an important efficiency improvement. However, apart from
improving the likelihood codes or waiting for the availability of
faster computers there is not much speed improvement to expect from an
MCMC approach, while
probably needed if one
wants to explore yet bigger and more complex models. On the
algorithmic side of the problem, some effort has been devoted recently
to the improvement of the likelihood codes, mainly by using clever
interpolation tricks (segmentation \citep{Fendt:2007p4993}, neural
networks \citep{Auld:2007p4956}) and by looking for improvements in
the MCMC algorithm \citep{hajian07,geman:geman:84,larsonetal07}. The
former \citep{Fendt:2007p4993,Auld:2007p4956} indeed provide some gain
in efficiency, but at the cost of a long pre-computation step for each
model. The latter improves on the natural inefficiency of the
Metropolis Algorithm but imposes some other requirements, like the
availability of cheap computation of the derivatives of the likelihood
\citep{hajian07}, or the knowledge of conditional probabilities of
some of the parameters \citep{geman:geman:84,larsonetal07}. Other (non
Markovian) Monte Carlo methods, such as nested sampling, have also
been proposed and applied recently to cosmological problems with some
success along with presenting their own problems
\citep{shawetal07,feroz:hobson:08,chopin:robert:08}.

On the hardware side, however, there is a route to speed improvement
that does not lie in quicker CPUs, but on the availability of cheap
multi-CPU computers and the standardization of clusters of
computers. This opportunity, however, is only partly opened to
MCMC. Indeed, there are two ways of parallelizing the parameter
exploration. First, by distributing the computation of the likelihood,
which is not always
possible and does not always lead to speed
improvement. Second, by running multiple chains in parallel. This last
option is the simplest, but is `forbidden' by the iterative nature of
the MCMC algorithm. More precisely, running parallel chains and mixing
them in the end to build a bigger chain sample is of course possible
(and can be advantageous in fully exploring the support of $\pi$), but
at the condition that each of the individual chains has converged. In
the absence of such a condition, significant biases in the sample can
be introduced. 
Determining convergence for each chain is inherently difficult in 
practice and has largely prevented more widespread use of the approach
\citep{rosenthal00}. Thus, for MCMC any speed improvements through
parallelization are difficult to achieve.

In this paper, we propose another sampling algorithm suitable for
cosmological applications, that is not based upon MCMC, and can be
parallelised. This novel algorithm, called Population Monte Carlo
(PMC) is an adaptive importance sampling technique, that
has been studied recently in the statistics literature
\citep{cappe:douc:guillin:marin:robert:2007}.  While this algorithm
solves some of the issues of MCMC in cosmology, the approach of course has a
different set of potential problems that we will analyse and discuss,
along with its advantages.

The paper is outlined as follows. In the next section, we provide a
brief introduction to the Bayesian approach, which we hope will give
the casual reader some important keys for further readings, and we
also discuss the challenges and issues involved with using either an MCMC
or an importance sampling algorithm for estimation.
We then describe details of the PMC approach.  In
Sect.~\ref{sec:simul}, we assess the performance of the PMC approach
using a simulated target density with features similar to cosmological
parameter posteriors, and provide a comparison to results obtained
using an MCMC approach. In Sect.~\ref{sec:cosmo}, we illustrate the
results from the PMC approach using actual data, consisting of CMB
anisotropies, supernovae of type Ia and weak cosmological lensing. We
conclude in Sect.~\ref{sec:discussion} with a discussion and an
outlook for further work.

%%%%%%%%%%%%%%%%%%%%%%%%%%%%%%%%%%%%%%%%%%%%%%%%%%%%%%%%%%%%
\section{Methods}
%%%%%%%%%%%%%%%%%%%%%%%%%%%%%%%%%%%%%%%%%%%%%%%%%%%%%%%%%%%%

%%%%%%%%%%%%%%%%%%%%%%%%%%%%%%%%%%%%%%%%%%%%%%%%%%%%%%%%%%%%
\subsection{Bayesian inference via simulation}
%%%%%%%%%%%%%%%%%%%%%%%%%%%%%%%%%%%%%%%%%%%%%%%%%%%%%%%%%%%%

A key feature of Bayesian inference is to provide a probabilistic
expression for the uncertainty regarding a parameter of interest $x$
by combining prior information along with information brought by the
data. Prior information, for example, could take the form of
information obtained from previous experiments which cannot readily be
incorporated into the current experiment or simply consist of a
feasible range.
The absence of prior information, however, is not restriction for the
use of Bayesian inference and estimation can still be regarded as
valid~\citep{robert:2001}. Information brought by the data and prior
information are entirely subsummed in the posterior probability
density function obtained, up to a normalization constant, by
\begin{equation}
  \pi(x) \propto \operatorname{likelihood}(\operatorname{data}|x) \times
  \operatorname{prior}(x).
  \label{eq:posterior}
\end{equation}
It is however generally difficult to handle the posterior
distribution, due to (a) the dimension of the parameter vector $x$,
and (b) the use of non-analytical likelihood functions.  For both of
these reasons, the normalizing constant missing from the right-hand
side of \eqref{eq:posterior} is usually not explicitly available.  A
practical solution to this difficulty is to replace the analytical
study of the posterior distribution with a simulation from this
distribution, since producing a sample from $\pi$ allows for a
straightforward approximation of all integrals related with $\pi$, due
to the Monte Carlo principle \citep{robert:casella:2004}. In short, if
$x_1,\ldots,x_N$ is a sample drawn from the distribution $\pi$ and $f$
denotes a function (with finite expectation under $\pi$), the
empirical average
\begin{equation}
  \frac{1}{N}\,\sum_{n=1}^N f(x_n)
  \label{eq:mcmc-estim}
\end{equation}
is a convergent estimator of the integral
\begin{equation}
  \pi(f) = \int f(x) \pi(x)\,\dd x,
  \label{some_integral}
\end{equation}
in the sense that the empirical mean \eqref{eq:mcmc-estim} converges
to $\pi(f)$ as $N$
grows to infinity.  Quantities of interest in a Bayesian analysis
typically include the posterior mean, for
which $f(x)=x$; the posterior
covariance
 matrix corresponding to
$f(x)= x x^\transp$;
and probability intervals, with $f(x) = \indicator_{S}(x)$, where $S$
is a domain of interest, and $\indicator_S(x)$ denotes the
indicator function which is equal to one if $x
\in S$ and zero otherwise.

%%%%%%%%%%%%%%%%%%%%%%%%%%%%%%%%%%%%%%%%%%%%%%%%%%%%%%%%%%%%
\subsection{Markov chain Monte Carlo methods}
%%%%%%%%%%%%%%%%%%%%%%%%%%%%%%%%%%%%%%%%%%%%%%%%%%%%%%%%%%%%

For most problems in practice, direct simulation from $\pi$ is not an option
and more sophisticated approximation techniques are necessary. One of
the standard approaches \citep{robert:casella:2004} to the simulation
of complex distributions is the class of Markov chain Monte Carlo
(MCMC) methods that rely on the production of a Markov chain $\{x_n\}$
having the target posterior distribution $\pi$ as limiting
distribution.

MCMC can be implemented with many Markovian proposal distributions but
the standard approach is the random walk Metropolis-Hastings
algorithm: given the current value $x_n$ of the chain, a new value
$x_\star$ is drawn from $\psi(x-x_n)$, where the so-called proposal
$\psi$ denotes a symmetric probability density function. The point
$x_\star$ is then accepted as $x_{n+1}$ with probability (also called
acceptance rate in this context)
\begin{equation}
  \min\left\{1, \frac{\pi(x_\star)}{\pi(x_n)}\right\},
  \label{eq:mh_acceptance}
\end{equation}
and otherwise, $x_{n+1}=x_n$. The algorithm is implemented as follows:\\

\smallskip

\begin{sffamily}
\hrule \centerline{{\bfseries \rule[-1.5mm]{0em}{5mm} Random walk Metropolis-Hastings
algorithm}} \hrule
\begin{itemize}
\item[Do:] Choose an arbitrary value of $x_1$.
\item[For] $n \geq 1$: \\
Generate $x_\star \sim \psi(x-x_{n})$ and $u\sim\operatorname{Uniform}(0,1)$.\\
Take
$$
x_{n+1} = \begin{cases}
x_\star & \text{if } u\le \pi(x_\star)\big/\pi(x_{n}), \\
x_{n} &\text{otherwise.}
\end{cases}
$$
\hrule
\end{itemize}
\end{sffamily}

\smallskip

While this algorithm is universal in that it applies to any choice of
posterior distribution $\pi$ and proposal $\psi$, its performance
highly depends on the choice of the proposal $\psi$ that has to be
properly tuned to match some characteristics of $\pi$. If the scale of
the proposal $\psi$ is too small, that is, if it takes many steps of
the random walk to explore the support of $\pi$, the algorithm will
require many iterations to converge and, in the most extreme cases,
will fail to converge in the sense that it will miss some relevant
part of the support of $\pi$ \citep{marin:mengersen:robert:2004}. If,
on the other hand, the scale of $\psi$ is too large, the algorithm may
also fail to adequately sample from $\pi$.  This time, the chain may
exhibit low acceptance rates and fail to generate a sufficiently diverse
sample, even with longer runs.  There exist monitors that assess the
convergence of such algorithms but they usually are conservative --
i.e., require a multiple of the number of necessary iterations -- and
partial -- i.e., only focus on a particular aspect of convergence or on
a special class of targets \citep{robert:casella:2004}. MCMC
algorithms are also notoriously delicate to calibrate on-line, both
from a theoretical point of view and from a practical perspective
\citep{haario:sacksman:tamminen:2001}.  For these approaches, often
called adaptive MCMC, some recommendations for the optimal scaling and
calibration schedule for various proposals in high dimensions have
been proposed \citep{roberts:rosenthal:01}, but this is still at an experimental stage.

%%%%%%%%%%%%%%%%%%%%%%%%%%%%%%%%%%%%%%%%%%%%%%%%%%%%%%%%%%%%
\subsection{Population Monte Carlo}
%%%%%%%%%%%%%%%%%%%%%%%%%%%%%%%%%%%%%%%%%%%%%%%%%%%%%%%%%%%%

Population Monte Carlo (PMC)
\citep{cappe:guillin:marin:robert:2003,cappe:douc:guillin:marin:robert:2007}
is an adaptive version of importance sampling
\citep{vonneumann:1951,rubinstein:1981} that produces a sequence
of samples (or populations) that are used in a sequential manner
to construct improved importance functions and improved estimations of the quantities of interest.

We recall that importance sampling is based on the fundamental
identity \citep{robert:casella:2004}
\begin{equation}
  \pi(f) = \int f(x) \pi(x) \,\dd x = \int f(x) \frac{\pi(x)}{q(x)} q(x) \,\dd x,
\label{eqn:isidentity}
\end{equation}
which holds for any probability density function $q$ with support including the support of
$\pi$ and any function $f$ for which the expectation $\pi(f)$ is finite. Hence, this approach
to approximating integrals
linked with complex distributions is also universal in that the above
identity always holds. If $x_1,\ldots,x_N$ are drawn
\emph{independently} from $q$,
\begin{equation}
  \hat \pi(f) = \frac{1}{N}\,\sum_{n=1}^N f(x_n) w_n; \quad w_n = \pi(x_n)/q(x_n),
  \label{eqn:isiapprox}
\end{equation}
provides a converging approximation to $\pi(f)$. In this context, $q$ is called
the importance function and $w_n$ are commonly referred
to as importance weights. For Bayesian inference, one cannot directly use
\eqref{eqn:isiapprox} as only the unnormalised version of $\pi$ (i.e., the
right-hand side of eq.~\ref{eq:posterior}) is available. Conveniently,
the self-normalised importance ratio
\begin{equation}
  \hat \pi_{\rm N}(f) = \sum_{n=1}^N f(x_n)\bar{w}_n,
  \label{eqn:isiapproxnorm}
\end{equation}
where the normalised importance weights are defined as
\begin{equation}
  \bar{w}_n = \frac{w_n}{\sum_{m=1}^{N}{w_m}},
  \label{wbar}
\end{equation}
is also a converging approximation to $\pi(f)$, independent of the
normalization of $\pi$.  For an importance function that is closely
matched to the target density, significant reductions in the variance
of the Monte Carlo estimates are possible in comparison to estimates
obtained using MCMC \citep{robert:casella:2004}. However, the importance sampling approach is equally prone to
poor performances as MCMC, in that the resulting converging
approximation may suffer from a large or even infinite variance if $q$ is not selected
in accordance with $\pi$. There is no universal importance function
and most of the research in this field aims at fitting the most
efficient importance functions for the problem at hand.

Population Monte Carlo offers a possible solution to
this difficulty through adaptivity: given the target posterior density $\pi$ up to a constant, PMC produces a sequence $q^t$ of importance functions $(t=1,\ldots,T)$ aimed at
approximating this very target. The first sample is produced by a
regular importance sampling scheme,
$x_1^{1},\ldots,x_N^{1}\sim q^1$, associated
with importance weights
\begin{equation}
  {w}_n^1 = \frac{\pi(x_n^{1})}{q^1(x_n^{1})};
  \quad n=1,\dots,N,
\end{equation}
and their normalised counterparts $\bar w_n^1$ (eq.~\ref{wbar}),
providing a first approximation to a sample from $\pi$. Moments of
$\pi$ can then be approximated to construct an updated importance
function $q^2$, etc.

The approximation can be measured in terms of the Kullback divergence
(also called Kullback-Leibler divergence or relative entropy) from the
target,
\begin{equation}
  K(\pi\|q^t) = \int \log\left(\frac{\pi(x)}{q^t(x)}\right) \pi(x) \dd x,
  \label{eqn:kdiv}
\end{equation}
and the density $q^t$ can be adjusted incrementally such that
$K(\pi\|q^t)$ is smaller than $K(\pi\|q^{t-1})$. The importance function should
be selected from a family of functions which is sufficiently large to allow for a
close match with $\pi$ but for which the minimization of \eqref{eqn:kdiv} is
computationally feasible. In \cite{cappe:douc:guillin:marin:robert:2007} the
authors propose to use mixture densities of the form
\begin{equation}
  q^t(x) = q(x;\alpha^t,\theta^t) = \sum_{d=1}^D \alpha^t_d \,
  \varphi(x;\theta^t_d)
  \label{eq:mixtureISdensity}
\end{equation}
where $\alpha^t = (\alpha_1^t, \ldots, \alpha_D^t)$ is a vector of adaptable
weights for the $D$ mixture components (with $\alpha_d^t > 0$ and
$\sum_{d=1}^D \alpha_d^t = 1$), and $\theta^t = (\theta_1^t, \ldots,
\theta_D^t$) is a vector of parameters which specify the components;
$\varphi$ is a parameterised probability density function, usually taken
to be multivariate Gaussian or Student-t (where the latter is to be
preferred in cases where it is suspected that the tails of the
posterior $\pi$ are indeed heavier than Gaussian tails). Given the
vast array of densities that can be approximated by mixtures, such an
importance function  provides considerable flexibility
to efficiently estimate a wide range of posteriors, including in this
case those found in cosmological settings.

The generic PMC algorithm then consists of the following:\\

\smallskip

\begin{sffamily}
\hrule \centerline{{\bfseries \rule[-1.5mm]{0em}{5mm} Population Monte Carlo algorithm}}
\hrule
\begin{itemize}
\item[Do:] Choose an importance function $q^1$.\\
Generate an independent sample $x_1^{1},\dots,x_N^{1}\sim q^1$.\\
Compute the importance weights ${w}_1^1,\dots,{w}_N^1$.
\item[For] $t \geq 1$:\\ Update the importance function to
  $q^{t+1}$, based on the previous weighted sample $(x_1^{t},
  w_1^{t}), \dots, (x_N^{t}, w_N^{t})$.\\ Generate independently
  $x_1^{t+1},\ldots,x_N^{t+1}\sim q^{t+1}$.\\ Compute the importance
  weights ${w}_1^{t+1},\ldots,{w}_N^{t+1}$.\\  \hrule
\end{itemize}
\end{sffamily}

\smallskip

Unlike for MCMC, in a PMC approach, the process can be interrupted at
any time as the sample produced at each iteration can be validly used
to approximate expectations under $\pi$ using self-normalised
importance sampling following \eqref{eqn:isiapproxnorm}. Further, sampling outputs from previous
iterations can be combined \citep{veach:guibas:95,AMIS:09}, and the sample size at each iteration does
not necessarily need to be fixed. Both of these properties of PMC can be
exploited to improve parameter estimates, either by increasing the
coverage of the importance function to the target density or
increasing the precision of the approximation for the integral of
interest.

Also note that an approximate sample from the \emph{target} density
can be obtained by sampling ($x_1^t,\dots,x_n^t$) with replacement,
using the normalised importance weights $\bar{w}_{n}^{t}$.  Although
this process induces extra Monte Carlo variation, there are a number
of methods available which considerably reduce the variation involved
(e.g.~residual sampling \citep{liu:chen:95} or systematic sampling
\citep{whitley94}).

%%%%%%%%%%%%%%%%%%%%%%%%%%%%%%%%%%%%%%%%%%%%%%%%%%%%%%%%%%%%
\subsubsection{Updating the importance function in the Gaussian case}
%%%%%%%%%%%%%%%%%%%%%%%%%%%%%%%%%%%%%%%%%%%%%%%%%%%%%%%%%%%%
\label{eq:gaussian_update}

In this section, we particularise the generic PMC algorithm to the
case where the importance function consists of a mixture of
$p$-dimensional Gaussian densities with mean $\mu_d$ and covariance
$\Sigma_d$,
\begin{align}
  \varphi(x;\mu_d,\Sigma_d) & = (2\pi)^{-p/2}|\Sigma_{d}|^{-1/2}
  \nonumber \\
  \times & \exp\left[-\frac{1}{2}(x-\mu_{d})^\transp
    \Sigma_{d}^{-1}(x-\mu_{d})\right].
  \label{eq:gaussian}
\end{align}

Using this importance function for the mixture model
\eqref{eq:mixtureISdensity}, we start the PMC algorithm by
arbitrarily fixing the mixture parameters $(\alpha^1,\mu^1,\Sigma^1)$,
and then sample independently from the resulting importance function
$q^1$ to obtain our initial sample $(x_{1}^1,\dots,x_{N}^1)$. From
this stage, updates of the parameters proceed recursively.

At iteration $t$, the importance weights associated with the
sample $(x_{1}^t,\dots,x_{N}^t)$ are given by
\begin{equation}
{w}_{n}^t=\frac{\pi(x_{n}^t)} {
\sum_{d=1}^{D}
\alpha_{d}^{t} \, \varphi(x_n^t;\mu_{d}^{t},\Sigma_d^t)}
\end{equation}
with normalised counterparts $\bar{w}^t_{n}$ given by eq.~(\ref{wbar}).
The parameters ($\alpha^t,\mu^t$ and $\Sigma^t$) of the importance function
are then updated according to
\begin{align}
& \alpha^{t+1}_{d} = \sum_{n=1}^N\bar{w}_{n}^t
  \, \rho_d(x_n^t;\alpha^{t},\mu^{t},\Sigma^t);
  \label{eq:update_alpha} \\
  &\mu^{t+1}_{d} = \frac{\sum_{n=1}^N\bar{w}_{n}^t x_{n}^t \, \rho_d(x_n^t;\alpha^{t},\mu^{t},\Sigma^t)}
{\alpha^{t+1}_{d}};
\label{eq:update_mu} \\
& \Sigma^{t+1}_{d} = \nonumber \\
& \quad \frac{\sum_{n=1}^N\bar{w}_{n}^t \,
  (x_{n}^t-\mu^{t+1}_{d})(x_{n}^t-\mu^{t+1}_{d})^\transp
  \rho_d(x_n^t;\alpha^{t},\mu^{t},\Sigma^t)}
{\alpha^{t+1}_{d}};
\label{eq:update_Sigma}
\end{align}
where
\begin{equation}
\rho_{d}(x;\alpha,\mu,\Sigma)= \frac{\alpha_d \,
  \varphi(x;\mu_{d},\Sigma_d)}{\sum_{d=1}^{D}\alpha_d \,
  \varphi(x;\mu_{d},\Sigma_d)}.
\end{equation}

Appendix \ref{sec:app-update-PMC} provides derivations of these
expressions and further details on the general approach, as well as
equations pertaining to the (more involved) case of mixtures of
multivariate Student-t distributions, which are used in the simulations
presented in Sect.~\ref{sec:simul}.

As discussed in Appendix \ref{sec:app-update-PMC}, the main theoretical appeal of this particular
update rule is that, as $N$ tends to infinity, the corresponding Kullback divergence
$K(\pi\|q^{t+1})$ is guaranteed to be less than $K(\pi\|q^{t})$.

%%%%%%%%%%%%%%%%%%%%%%%%%%%%%%%%%%%%%%%%%%%%%%%%%%%%%%%%%%%%
\subsubsection{Monitoring convergence}
\label{sec:PMC_monitor}
%%%%%%%%%%%%%%%%%%%%%%%%%%%%%%%%%%%%%%%%%%%%%%%%%%%%%%%%%%%%

The above update process can be repeated a number of times, and although there
is no need for a formal stopping rule, some measures of performance against the target
density can be used as a guide. As the objective of importance function adaptations is to
minimise the Kullback divergence between the target density and the importance function,
we can stop the process when further adaptations do not result in significant
improvements in $K(\pi\|q^t)$. To this end, it can be shown that $\exp[-K(\pi\|q^t)]$ may be
estimated by the \emph{normalised perplexity}
\begin{equation}
  p = \exp(H_{\rm N}^t)/N,
  \label{perplexity}
\end{equation}
where
\begin{equation}
  H_{\rm N}^t = -\sum_{n=1}^N \bar{w}_{n}^t\log\bar{w}_{n}^t
  \label{shannon-entropy}
\end{equation}
is the Shannon entropy of the normalised weights, a frequently used
measure of the quality of an importance sample. Thus,
minimization of the Kullback divergence can be approximately connected
with the maximization of the perplexity (\ref{perplexity}). Values of this criterion
close to 1 will therefore indicate good agreement between the
importance function and the target density.

Another frequently used criterion for importance sampling is the
so-called \emph{effective sample size }(ESS),
\begin{equation}
  \operatorname{ESS}_{\rm N}^t = \left( \sum_{n=1}^{N}
  \left\{\bar{w}_n^t\right\}^2 \right)^{-1},
  \label{ess}
\end{equation}
which lies in the interval $[1; N]$ and can be interpreted as the
number of sample points with non-zero weight
\citep{liu:chen:1995}. Both measures (\ref{perplexity}, \ref{ess})
are interconnected, as an importance function which is close to the
target density will have both a high normalised perplexity and a
relatively large number of points with non-zero weight, compared to an
ill-fitting importance function.  Given a real-valued function $f$ of
interest one can also estimate the asymptotic variance of the
self-normalised importance sampling estimator $\hat{\pi}_{\rm
  N}^t(f)=\sum_{n=1}^{N}\bar{w}_{n}^t f(x_n^t)$
(cf.~eq.~\ref{eqn:isiapproxnorm}) using the importance sample itself,
as
\begin{equation}
  N \sum_{n=1}^N \left\{\bar{w}_n^t
  \left(f(x_n^t)-\hat{\pi}_N^t(f)\right) \right\}^2.
  \label{eqn:varf}
\end{equation}
Beware that this formula (which is derived from theorem 2
of \cite{RePEc:ecm:emetrp:v:57:y:1989:i:6:p:1317-39}) 
 is only valid with normalised weights,
and that it is a variance conditional on the current importance
  proposal $q^{t}$, i.e.~it does not take into account the
  adaptation. This measure can
be related to the so-called \emph{integrated autocorrelation time
  (IAT)} used for Markov chain Monte Carlo simulations, which, in this
case, takes into consideration the level of autocorrelation present in
the chain
\citep{robert:casella:2004,roberts:rosenthal:01,2005MNRAS.356..925D}.

%%%%%%%%%%%%%%%%%%%%%%%%%%%%%%%%%%%%%%%%%%%%%%%%%%%%%%%%%%%%
\subsubsection{A first illustration}
\label{sec:illustration}
%%%%%%%%%%%%%%%%%%%%%%%%%%%%%%%%%%%%%%%%%%%%%%%%%%%%%%%%%%%%
\begin{figure}[t!]
  \centering
  \includegraphics[width=0.48\textwidth]{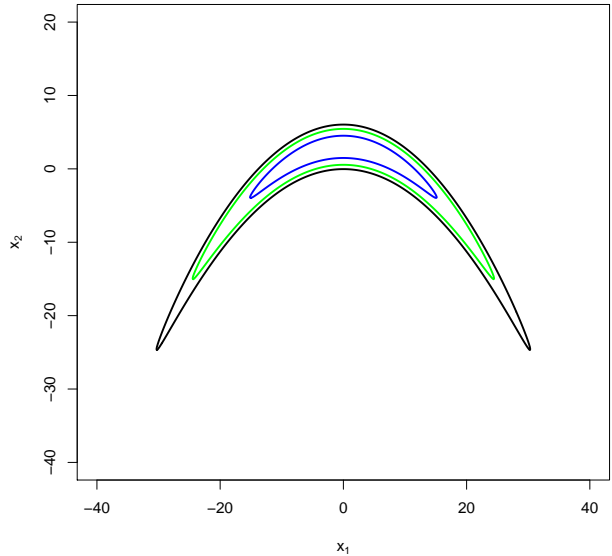} \\
  \caption{Test target density on the
    ($x_1,x_2$) plane.  Contours represent the 68.3$\%$ (blue), 95$\%$
    (green) and 99\% (black) confidence regions.  }
  \label{fig:contourd2b}
\end{figure}

\begin{figure}
\centering
\includegraphics[width=0.48\textwidth]{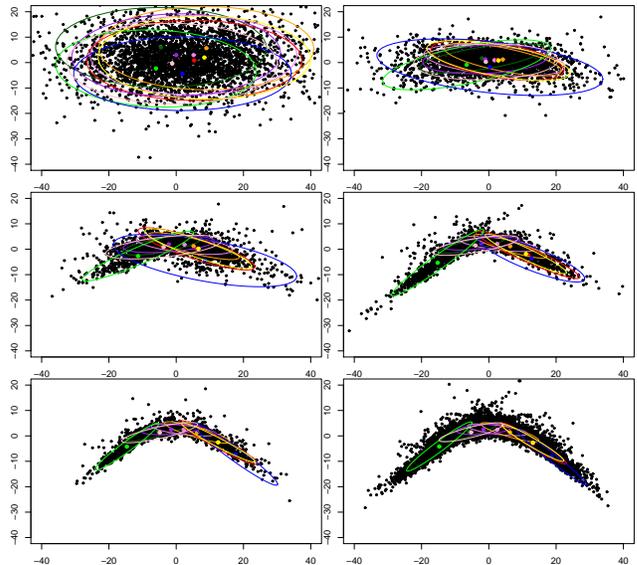}
\caption{\small Evolution of the importance function for the target density (see Fig.~\ref{fig:contourd2b}) over 11 iterations of 10k points for $x_1$ (horizontal axis) and $x_2$ (vertical axis), except for the last iteration (11) which is a sample of 100k points. Iterations 1 (top left) to 11 (bottom right) from left to right with every second iteration shown (i.e 1,3,5,7,9,11). Colours indicate the mixture components with mean of each component indicated by coloured dots and approximate $95\%$ confidence regions for the sample of points from each component by coloured ellipses. Every 3rd sample point from the importance function is plotted.
}
\label{fig:res-d10-b003-evolution}
\end{figure}

To illustrate the PMC approach, we explore a banana-like target density 
presented Fig.~\ref{fig:contourd2b}. The same target distribution will be 
studied in greater depth in the next section. 
The results of the first eleven (11) iterations of the PMC algorithm using 
a mixture of Student-t densities are shown Fig.~\ref{fig:res-d10-b003-evolution} 
(see also Appendix \ref{sec:app-update-PMC} for details of the update procedure).

\begin{figure}
\centering
\includegraphics[width=0.48\textwidth]{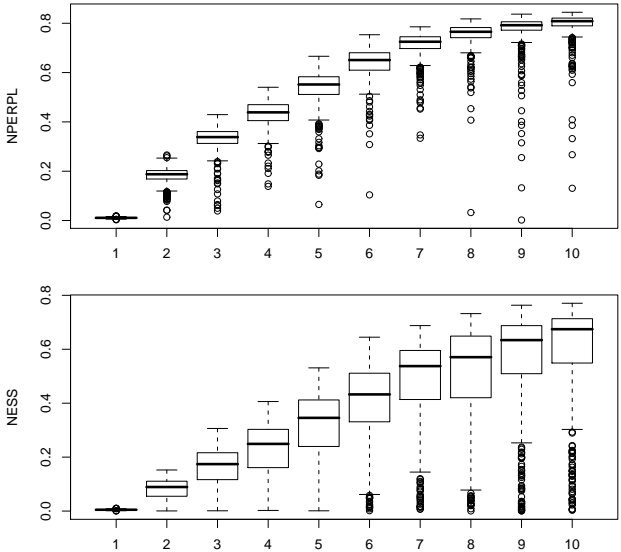}
\caption{Normalised perplexity (top panel) and normalised effective sample size
  (ESS/$N$) (bottom panel) estimates for the first 10 iterations of
  PMC (represented in Fig.~\ref{fig:res-d10-b003-evolution}) over 500
  simulation runs. The distributions are shown as whisker plots: the
  thick horizontal line represents the median; the box shows the
  interquartile range (IQR), containing 50\% of the points; the
  whiskers indicate the interval $1.5\times$IQR from either Q1 (lower)
  or Q3 (upper); points outside the interval $[Q1, Q3]$ (outliers) are
  represented as circles.}
\label{fig:essfig}
\end{figure}

While this target density shows slightly more 
pronounced curvature for an example of a posterior density in
practice, it serves to illustrate the process of adaptation of the
importance function. The initial importance function $q^1$ is a
mixture of multivariate Student-t's, consisting of nine components
placed around the centre of the range for the first two variables,
each with a relatively large variance (for the first dimension $=200$, second $=50$) and degrees of freedom
$\nu=9$. The different coloured circles in Fig.~\ref{fig:contourd2b}
indicate the location of the component means, and the circle size is
proportional to the weight $\alpha_d$ associated with the component. At
the fourth iteration ($t=4$), we see that the importance function
starts to resemble the shape of the target density, with components
becoming more separated and moving into the tails of the target.  By
the sixth iteration ($t=6$) the importance function has further
adapted to the shape of the target banana density. For this target
density and importance function,  Fig.~\ref{fig:essfig} presents
estimates of the normalised perplexity and normalised effective sample
size (ESS/$N$) for the first 10 iterations over 500 simulation runs.
As shown, the estimates of the normalised perplexity improve rapidly
from approximately 0.14 for the second importance function (Iteration
2) to approximately 0.81 for the last importance function (Iteration
10), with a similar increase in estimates of the normalised effective sample size (ESS/$N$, increasing from 0.10 to 0.60).
For this importance function and target density, the normalised perplexity starts to level off after the 10th iteration (around $0.82$), indicating that there is no need for further adaptation of the proposal density. As mentioned previously however, in general, one does not need to observe the convergence of the proposal (as for MCMC) in order to stop the sampling process.

An important consideration, and a choice that needs to be made at the start of
the algorithm are the parameter values $\alpha^1$, $\mu^1$ and
$\Sigma^1$ for the initial importance function $q^1$, including the
degrees of freedom $\nu$ in the case of the Student-t mixture, and the
sample size $N$. A poor initial importance function, such as one that
is tightly centred around only one mode in the case of a multimodal
posterior or a narrow importance function with light tails, may take a
long time to adapt or may miss important parts of the posterior.  For
importance sampling the choice of $q$ requires both fat tails and a
reasonable match between $q$ and the target $\pi$ in regions of high
density. Such an importance function can be more easily constructed in
the presence of a well informed guess about the parameters and
possibly the shape of the posterior density.
Sample size considerations
also play an important role -- smaller samples can adapt quite
quickly with less computational time but may provide less reliable
information about the posterior density relative to larger samples.
Such considerations are important as we look at posterior densities of
increasingly high dimensions, and thus we can expect to take a
larger sample size as the dimension of the problem increases. We will
discuss these issues further in the context of simulated and
actual data, and also in Sect.~\ref{sec:discussion}.

%%%%%%%%%%%%%%%%%%%%%%%%%%%%%%%%%%%%%%%%%%%%%%%%%%%%%%%%%%%%
\section{Simulations}
\label{sec:simul}
%%%%%%%%%%%%%%%%%%%%%%%%%%%%%%%%%%%%%%%%%%%%%%%%%%%%%%%%%%%%

In this section, we test the performance of PMC using simulated data, and
compare the results to an adaptive MCMC procedure.

\setcounter{subsubsection}{0}
\subsubsection{Target density}

In order to provide a good test for both approaches we use the target
density considered in~\citep{haario:sacksman:tamminen:2001}, which is
difficult to explore but which also provides a realistic scenario for
many problems encountered in cosmology. The target density is based on
a centred $p$-multivariate normal, $x \sim {\cal N}_p(0,\Sigma)$ with
covariance $\Sigma=\textup{diag}(\sigma_1^2,1,\dots,1)$, which is
slightly twisted by changing the second co-ordinate $x_2$ to $x_{2}+b
(x_{1}^2- \sigma_1^2)$. Other co-ordinates are unchanged.  We obtain a
twisted density which is centered with uncorrelated components.  Since
the Jacobian of twist is equal to~1, the target density is:
\begin{equation}
  (x_1,x_2+b (x_1^2- \sigma_1^2),x_3,\dots,x_{p}) \sim {\cal N}_p(0,\Sigma) \,.
\end{equation}
For the target density that we will consider, we set $p=10$, $\sigma_1^2 =
100$, $b=0.03$, which results in a banana-shaped density in the first
two dimensions (see Fig.~\ref{fig:contourd2b}).

For the target density considered, interest is in how well PMC and MCMC are
able to approximate the tails of the target.  Whilst the curvature present in
the first two dimensions of this target density is slightly more pronounced
than what is typically seen in practice for cosmology it serves to highlight
the difficulties faced by both PMC and MCMC in covering the parameter space. In
particular, little accurate information is available in order to guide
the choice of importance function (for PMC) and proposal distribution (for
MCMC) and so both approaches are forced to learn about the parameter space.

\subsubsection{Test run proposal for PMC}
\label{sec:propPMC}

For PMC, and in the absence of any detailed \emph{a priori} information about
the target density, except the possible range for each variable, we have chosen
the first importance function to be a mixture of multivariate Student-t
distributions with components displaced randomly in different directions
slightly away from the centre of the range for each variable: the mean of the
components is drawn from a $p$-multivariate Gaussian with mean $0$ and
covariance equal to $\Sigma_0/5$ where $\Sigma_0$ is some
positive-definite matrix; the variance for components was chosen to be
$\Sigma_0$. We choose a mixture of $9$ components of Student-t distribution
with $\nu=9$ degrees of freedom; and $\Sigma_0$ is a diagonal matrix with
diagonal entries $(200, 50, 4, \cdots, 4)$. This choice of $(\nu, \Sigma_0)$
ensures adequate coverage, albeit somewhat overdispersed, of the feasible
parameter region. In this simulated example, Student-t distributions are preferable
to Gaussian distributions because the range of the variables is unbounded (in contrast to the
cosmology examples to be discussed in Sect.~\ref{sec:cosmo}).

A representation of the first importance function for the first two dimensions
is shown in the top left-hand box in Fig.~\ref{fig:res-d10-b003-evolution}, with
a typical evolution over the next few iterations in the other panels.  In pilot
runs of various importance functions against the target density, the best
fitting importance function required at least seven components in order to
adequately represent the coverage of the entire density.

For PMC, an important issue is the sample size for each iteration.  A poor
initial importance function with a relatively small sample size will take a
long time to adapt or it may even be unable to recover sufficiently to provide
reasonable parameter estimates.  Such problems are more likely to occur as the
dimension of the parameter space increases, the so-called curse of
dimensionality. For the simulation exercise each iteration is based on a sample
of $10$k points. To prevent numerical instabilities in the updating of the parameters, components with a very small weight ($< 0.002$) or containing less than $20$ sample points are discarded.

\subsubsection{Test run proposal for MCMC}
\label{sec:propMCMC}

As little information is available for the target density, an adaptive
MCMC approach is used which can allow for faster learning of the target density
than using either independent or non-adaptive random walk
proposals~\cite{robertsetal97}.  For MCMC, the proposal distribution is
a centred Gaussian with covariance matrix which is updated along the
iterations.  An important choice for adaptive MCMC concerns the scaling of the
proposal and the rate of adaptation.  There has been much research on
this~\cite{robertsetal97,fort08}, and a common choice for the covariance of the Gaussian proposal is to consider $c \, {\Sigma}_n$ where
${\Sigma}_n$ is an estimate of the covariance of the target density, at update
$n$.  The choice $c = 2.38^2/p$ is considered to be optimal when the chain is
in its stationary regime, and for target densities that have a product
form~\cite{robertsetal97}. However for the target density we consider
this does not hold: the first two components are not independent despite being
uncorrelated and dependence is not linear but quadratic. However, with no other
theoretical results to follow we start with a scaling factor of that form and
for the simulation results to follow assess the effect on convergence and
results using alternative values.  We update the covariance matrix by the
recursive formula
\begin{equation}
{\Sigma}_n=(1-a_n){\Sigma}_{n-1}+a_{n} S_n
\end{equation}
where ${\Sigma}_{n-1}$ is the sample covariance of the previous update, and
${S}_n$ is the covariance of the sampled estimates from the previous update to
the current iteration. The value of $a_n$ is $1/n^k$ with $k$ chosen suitably
to allow for a cooling of the update, which is a necessary condition to ensure
convergence of this adaptive MCMC to the target density as well as convergence
of the empirical averages~\cite{robertsrosenthal07,fort08}.

In pilot runs, we explored the effect of this schedule for various values of
$(k,c) $ in $(0,1) \times (0, 2.38^2/p) $ and we observe that the choice of
$(k,c)$ plays a role on the time to convergence (for the estimation of the
quantities of interest, see below) and on the acceptance rate of the chain.

To ensure a fair comparison with PMC, we start the chain at a random point
drawn from the same Gaussian distribution as for PMC (i.e
$\mathcal{N}_d(0,\Sigma_0/5)$, using the same values for $\Sigma_0$ as used for
PMC).  We also explored in pilot runs the role of the initial value of the
chain: despite it being known that MCMC is sensitive to the choice of the initial
position of the chain - which has no real counterpart in PMC - this hasn't been
found to have a major impact on performance (for reasonable choice of the
initial value at least) in this particular study. We also fixed the update
schedule to be every $10$k points and we assessed the effect on the results
from using less or more points before updating. For the simulation results to
follow, $(k,c)$ has been set to $(0.5, 2.38^2/p)$ which ensures convergence
after the burn-in period (see Sect.~\ref{sec:test_runs}), and a mean
acceptance rate at convergence of about $10\%$. The proposal distribution is
updated for the entire length of the chain and is not stopped after the burn-in
period.

\subsection{Test runs}
\label{sec:test_runs}

For PMC and the proposal outlined, the perplexity appeared to level off at
around the $10$th iteration, so for the results to follow for PMC we ran the
PMC algorithm for $10$ iterations ($10$k points per iteration) and used a
final draw of $100$k points. To assess MCMC for the same number of points we
used a chain length of $200$k points with a burn-in of $100$k points.  Results
for both approaches at successive intervals before $200$k points are also
provided.  To assess the performance of the approaches, each simulation was
replicated $500$ times.

%%%%%%%%%%%%%%%%%%%%%%%%%%%%%%%%%%%%%%%%%%%%%%%%%%%%%%%%%%%%
\subsection{Results of the simulations}
\label{sec:results_sim}
%%%%%%%%%%%%%%%%%%%%%%%%%%%%%%%%%%%%%%%%%%%%%%%%%%%%%%%%%%%%

For the results of the simulations, we are interested in both the mean
estimates of the parameters (in particular for $x_1$ and $x_2$) and also
estimates of the confidence region coverage ($68.3\%$ and $95\%$) which will
provide an indication of how well both approaches are covering the tails of the
target density.
For each run $r=1,\ldots,500$, we provide the results for various functions $f$
of interest:
\begin{align*}
  f_a(\mathbf{x}) &= x_1 \\
  f_b(\mathbf{x}) &= x_2 \\
  f_c(\mathbf{x}) &= \mathbf{1}_{68.3} (\mathbf{x})  \\
  f_d(\mathbf{x}) &= \mathbf{1}_{95} (\mathbf{x})  \\
  f_e(\mathbf{x}) &= \mathbf{1}_{68.3} (x_1, x_2)  \\
  f_g(\mathbf{x}) &= \mathbf{1}_{95} (x_1, x_2) \\
  f_h(\mathbf{x}) &= \mathbf{1}_{68.3} (x_1)  \\
  f_i(\mathbf{x}) &= \mathbf{1}_{95} (x_1)
\end{align*}
We note here $\mathbf{1}_{q}$ as the indicator function for the $q\%$ region.
$f_h$ and $f_i$ are indicators only for the first dimension, while $f_e$ and
$f_g$ are dealing with the first 2. In all cases, the remaining dimensions are
marginalised over.

\begin{figure}[htp!]
\centering
\includegraphics[width=0.48\textwidth]{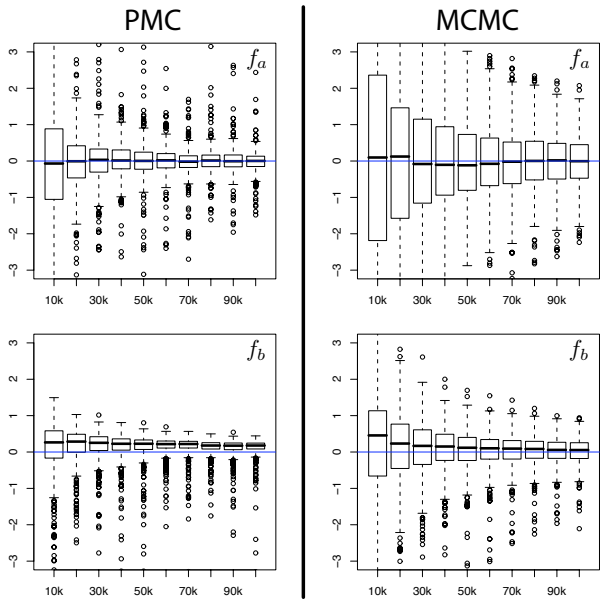}
\caption{\small Evolution of $\pi(f_a)$ (top panels) and $\pi(f_b)$
  (bottom panels) from 10k points to 100k points for both PMC (left
  panels) and MCMC (right panels). See
    Fig.~\ref{fig:essfig} for details about the whisker plot
    representation.}
\label{fig:res-d10-b003-evolution-comp}
\end{figure}

Table~\ref{tab:results} shows the results for estimation of $\pi(f)$
for functions $f_a$ and $f_b$ ($\bar{x}_1$ and $\bar{x}_2$
respectively).  The results provided show the mean and standard
deviation of estimates calculated over 500 runs. Although the
performance is quite similar for both methods, PMC does display a
two-fold reduction in standard deviation compared to MCMC for both
functions. A closer look at the results reveals that for $\pi(f_{a})$
the empirical distributions of the estimates (see
Figure~\ref{fig:res-d10-b003-evolution-comp}) are quite similar for
both methods, except for the variance which is much reduced for
PMC. For $\pi(f_{b})$, on the other hand, the empirical distribution
of the estimates for PMC are quite skewed, resulting in a slight
positive bias for the majority of the runs (second panel of
Figure~\ref{fig:res-d10-b003-evolution-comp}). The difference between
$\pi(f_{a})$ and $\pi(f_{b})$ can be explained from
Figure~\ref{fig:contourd2b} which shows that failure to visit
sufficiently the downward low-probability tails does indeed imply a
positively biased estimates for the mean of the second component. PMC
does appear to be more sensitive to this issue than MCMC, despite the
fact that the estimates for MCMC display a larger overall variability.

\begin{table}[b!]
\caption{Results of the simulations for the $10$-dimensional banana shaped target density over 500 runs for both PMC and MCMC}
\label{tab:results}
\input{inputs/table_simresults}
\end{table}

\begin{figure}[t!]
  \centering
  \includegraphics[width=0.48\textwidth]{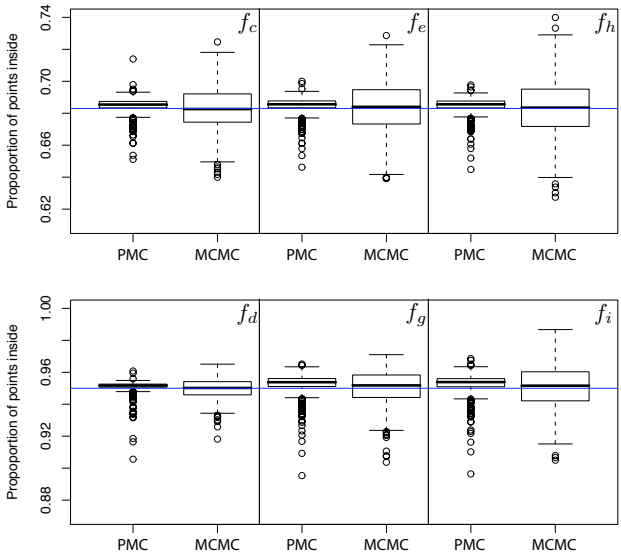}
  \caption{Results showing
    the distributions of the PMC and the MCMC estimates $ \pi(f)$ for (top)
    $f=f_c, f_e, f_h$ and (bottom) $f=f_d, f_g,f_i$ (in this order, left to
    right).  All estimates are based on 500 simulation runs. See
    Fig.~\ref{fig:essfig} for details about the whisker plot
    representation.}
    \label{fig:res-d10-b003-CIs}
\end{figure}

Figure~\ref{fig:res-d10-b003-CIs} provides the results for the confidence
region coverage.  To depict the variability of the data, the results are
displayed by using whisker plots. The results from both PMC and MCMC against all of
the performance measures are similar, with both showing good coverage of the
target density. The distribution of this estimator is again more skewed for PMC than
it is for MCMC, particularly for the 95\% regions in the bottom panel of
Figure~\ref{fig:res-d10-b003-CIs}. Nevertheless, the variability of the estimates obtained
with PMC also is significantly reduced compared to MCMC.

Figure~\ref{fig:res-d10-b003-evolution-comp} shows the evolution of the results
for $\pi(f_a)$ and $\pi(f_b)$ from $10$k points to $100$k points for both PMC
(left panels) and MCMC (right panels).
The results from Fig.~\ref{fig:res-d10-b003-evolution-comp}, in general,
highlight the reduction in variance of the Monte Carlo estimates for PMC in
comparison to MCMC. In particular, it is interesting to note that the variance
of the estimates, either for $\pi(f_a)$ or $\pi(f_b)$ for 100k posterior evaluations under MCMC
are comparable to estimates obtained using PMC at the second iteration (20k
points).

Simulating from this target distribution is a challenging problem for both
methods.  In particular, the use of a vague initial importance function in a
multidimensional space represents a challenge to PMC and it has been observed
that the importance function takes some time to properly adapt to the target
density (about 10 iterations). The choice of the initial importance function
in PMC is more crucial than is the choice of the initial proposal distribution
in adaptive MCMC. Although different variations for updating the covariance matrix for the MCMC
approach are possible we did not see a significant improvement in the results
presented from using alternative covariance structures. For most of the
simulation results, the proposal covariance matrix was observed to adapt
relatively quickly to the true covariance matrix. Changes to the covariance
structure considered included changes to the update frequency, the starting
proposal $\Sigma_0$, the scaling of the proposal (value of $c$) and adaptation
of the covariance (value
of $k$). 
Hence, the PMC approach may require more precise \emph{a priori} knowledge
of the target density than MCMC.

In the next section, we apply the PMC approach to typical cosmological
examples, and provide results in comparison to MCMC.

%%%%%%%%%%%%%%%%%%%%%%%%%%%%%%%%%%%%%%%%%%%%%%%%%%%%%%%%%%%%
\section{Application to cosmology}
\label{sec:cosmo}
%%%%%%%%%%%%%%%%%%%%%%%%%%%%%%%%%%%%%%%%%%%%%%%%%%%%%%%%%%%%

We apply our new adaptive importance sampling method to the posterior
of cosmological parameters. Flat CDM models with either a cosmological
constant ($\Lambda$CDM) or a constant dark-energy equation-of-state
parameter ($w$CDM) are explored and tested with recent observational
data of CMB anisotropies, supernovae type Ia and cosmic shear, as
described in the next section.

The three data sets and
likelihood functions used here are the same as in \cite{KB08};
the CMB measurements and likelihood are based on the five-year WMAP data release
\cite{WMAP5-Hinshaw08}, the SNIa data set is the first-year SNLS survey
\cite{2006A&A...447...31A}, while the cosmic shear is from 
the CFHTLS-Wide third release \cite[T0003,][]{FSHK08}.  The
results presented in the following sections can be compared to the
MCMC analysis in \cite{KB08}.

%%%%%%%%%%%%%%%%%%%%%%%%%%%%%%%%%%%%%%%%%%%%%%%%%%%%%%%%%%%%
\subsection{Data sets}
%%%%%%%%%%%%%%%%%%%%%%%%%%%%%%%%%%%%%%%%%%%%%%%%%%%%%%%%%%%%

%%%%%%%%%%%%%%%%%%%%%%%%%%%%%%%%%%%%%%%%%%%%%%%%%%%%%%%%%%%%
\subsubsection{CMB}
%%%%%%%%%%%%%%%%%%%%%%%%%%%%%%%%%%%%%%%%%%%%%%%%%%%%%%%%%%%%

To obtain theoretical predictions of the CMB temperature and
polarization power- and cross-spectra we use the publicly available
package CAMB \citep{Lewis:1999bs}.  The likelihood is calculated using
the public WMAP5 code \citep{WMAP5-Dunkley08}.

The WMAP5 likelihood takes as input the TT,TE,EE and BB theoretical
power spectra calculated by CAMB, and returns a likelihood computed
from a sum of different parts. It computes a pixel-based Gaussian
likelihood based on template-cleaned maps \citep{WMAP5-Gold08} and
their associated inverse covariance matrices (see \citet{WMAP3-Page07}
for details) at large angular scales ($\ell \le 32$ for TT, $\ell \le
23$ for TE, EE and BB). At small angular scales, it computes an
approximate likelihood based on pseudo-spectra and their associated
covariance for TT and TE \citep{WMAP3-Hinshaw07}, based respectively
on the (Q,V) and (V,W) channel pairs for TE and TT.

In addition, the likelihood computation takes into account analytic
marginalisations on nuisance parameters such as the beam transfer
function and point-sources uncertainties
\citep{WMAP3-Hinshaw07,WMAP5-Nolta08}.  We ignore corrections due to
SZ
and impose a larger (flat) prior on the Hubble constant.  Indeed,
CMB data alone exhibits a degeneracy between the Hubble
constant and e.g.~the cosmological constant \citep{Efstathiou99} which
is removed by adding other cosmological probes.

The acoustic oscillation peaks in the CMB anisotropy spectrum are a
standard ruler at a redshift of about $z=1100$. CMB therefore
measures the angular diameter distance to that redshift which depends
mainly on the total matter-energy density ($\Omegam + \Omegade$) and
weakly on the Hubble constant $h$. The overall anisotropy amplitude is
determined by the large-scale normalization $\Delta_R^2$. The relative
height of the peaks is sensitive to the baryonic and dark matter
densities. On large scales, secondary anisotropies are generated at late
times ($z \lsim 20$) due to reionisation, which is parametrised by the
optical depth $\tau$, and the integrated Sachs-Wolfe (ISW) effect,
which is a probe of $\Omegade$.

%%%%%%%%%%%%%%%%%%%%%%%%%%%%%%%%%%%%%%%%%%%%%%%%%%%%%%%%%%%%
\subsubsection{SNIa}
%%%%%%%%%%%%%%%%%%%%%%%%%%%%%%%%%%%%%%%%%%%%%%%%%%%%%%%%%%%%

The SNLS data set is described in detail in
\cite{2006A&A...447...31A}. We use their results from the SNIa
light-curve fits which, for each supernova, provides the rest-frame
$B$-band magnitude $m_B^*$, the shape or stretch parameter $s$ and
the colour $c$. We use the standard likelihood analysis described in
\cite{KB08}, adopted from \cite{2006A&A...447...31A}.

Under the assumption that supernovae of type Ia are standard candles
we can fit the luminosity distance to the SNIa data. The luminosity
distance is a function of $\Omegam$, $\Omegade$ and $w$. Three
additional parameters are the universal SNIa magnitude $M$ and the
linear response parameters to stretch and colour, $\alpha$ and $\beta$,
respectively. Those three parameters are specific to our choice of
distance estimator, and can be regarded as nuisance parameters.
The Hubble constant $h$ is integrated into the parameter $M$, so there
is no explicit dependence on $h$ in the SNIa posterior.

%%%%%%%%%%%%%%%%%%%%%%%%%%%%%%%%%%%%%%%%%%%%%%%%%%%%%%%%%%%%
\subsubsection{Cosmic shear}
%%%%%%%%%%%%%%%%%%%%%%%%%%%%%%%%%%%%%%%%%%%%%%%%%%%%%%%%%%%%

The CFHTLS-Wide $3^{\rm rd}$ year data release (T0003), the data and
weak lensing analysis as well as cosmological results are described in
\cite{FSHK08}.
As in \cite{FSHK08} we use the aperture-mass dispersion between 2 and
230 arc minutes as second-order lensing observable
\citep{1998MNRAS.296..873S}. We assume a multivariate Gaussian
likelihood function and take into account the correlation between
angular scales. The theoretical aperture-mass dispersion is obtained
by non-linear models of the large-scale structure
\cite{2003MNRAS.341.1311S}.

The galaxy redshift distribution is obtained by using the CFHTLS-Deep
redshift distribution \cite{2006A&A...457..841I} and rescaling it
according to the $i_{AB}$ magnitude distribution of CFHTLS-Wide
galaxies. We fit the resulting histogram with eq.~(14) from
\cite{FSHK08}, introducing the three fit parameters $a, b, c$. The
histogram data is modeled as multivariate, uncorrelated Gaussian, the
corresponding likelihood is included, independent of the lensing
likelihood, in the analysis.

Weak gravitational lensing by the large-scale structure is sensitive
to the angular diameter distance and the amount of structure in the
Universe.  It is an important probe to measure the normalisation
$\sigma_8$ on small scales. With the current data, this parameter is
however largely degenerate with $\Omegam$. This degeneracy is likely
to be lifted by future surveys which will include the measurement of
higher-order statistics \cite{2004MNRAS.348..897T,KS05} and shear
tomography \cite{1999ApJ...522L..21H}. In particular from the latter a
great improvement on the determination of $w$ is to be expected, a
parameter which is only weakly constrained by lensing up to now
\cite{CFHTLSwide, JBBD06}.

%%%%%%%%%%%%%%%%%%%%%%%%%%%%%%%%%%%%%%%%%%%%%%%%%%%%%%%%%%%%
\subsection{Cosmological parameter and priors}
%%%%%%%%%%%%%%%%%%%%%%%%%%%%%%%%%%%%%%%%%%%%%%%%%%%%%%%%%%%%

We sample a hypercube in parameter space which corresponds to flat
priors for all parameters, see Table \ref{tab:parameters} for more
details. Additional priors exist, both in explicit and implicit form,
which represent regions of parameter space which are unphysical or
where numerical fitting formulae break down. For example, we exclude
extremely high baryon fractions ($\Omegab > 0.75\Omegam$) because of
numerical problems in the computation of the transfer
function. Further, for very low values of both $\Omegam$ and
$\sigma_8$ the pivot scale for the non-linear power spectrum is
outside the allowed range. Very rarely, the calculation of the
likelihood for individual points in parameter space is unsuccessful
because of numerical errors or limitations of the likelihood code. Since
these points cannot be taken into account, a pragmatic solution is to formally modify the
prior to exclude those points. Note that these rare cases occur mainly
in regions of very low likelihood.

\begin{table*}
  \caption{Parameters for the cosmology likelihood. C=CMB, S=SNIa, L=lensing.}
  \label{tab:parameters}
  \begin{tabular}{lllllll} \hline\hline
    Symbol & Description & Minimum & Maximum &
    \multicolumn{3}{c}{Experiment} \\ \hline
    $\Omegab$ & Baryon density & 0.01 & 0.1 & C & & L \\
    $\Omegam$ & Total matter density & 0.01 & 1.2 & C & S & L\\
    $w$       & Dark-energy eq.~of state & -3.0 & 0.5 & C & S & L \\
    $n_{\rm s}$ & Primordial spectral index & 0.7 & 1.4 & C & & L \\
    $\Delta_R^2$ & Normalization (large scales) & & & C & & \\
    $\sigma_8$ & Normalization (small scales)\footnote{For WMAP5, $\sigma_8$ is a deduced quantity
      that depends on the other parameters} & & & C & & L \\
    $h$        & Hubble constant & & & C & & L \\
    $\tau$    & Optical depth & & & C & & \\ \hline
    $M$        & Absolute SNIa magnitude & & & & S & \\
    $\alpha$   & Colour response          & & & & S & \\
    $\beta$    & Stretch response        & & & & S & \\
    $a$        &                     & & & & & L \\
    $b$        & galaxy $z$-distribution fit & & & & & L \\
    $c$        &                     & & & & & L \\ \hline
    \hline
  \end{tabular}
\end{table*}

%%%%%%%%%%%%%%%%%%%%%%%%%%%%%%%%%%%%%%%%%%%%%%%%%%%%%%%%%%%%
\subsection{Initial choice of the importance function}
%%%%%%%%%%%%%%%%%%%%%%%%%%%%%%%%%%%%%%%%%%%%%%%%%%%%%%%%%%%%

As described earlier in Sect.~\ref{sec:illustration}, it is important
to have a good guess for the initial importance function. In all cases
considered here, we rely on an estimate of the maximum
likelihood point and the Hessian at that point (Fisher matrix) to build our initial proposals. 
We use
the conjugate-gradient approach \cite{nr} to find the maximum
likelihood point at which to calculate the Fisher matrix $F$ using the
theoretical model. We construct a mixture model consisting of $D$
Gaussian components. Student-t mixtures with small degrees of freedom
were tested and turned out to be a bad
approximation to the posterior under study, resulting in low
perplexities. Each mixture component is shifted randomly from the
maximum by a small amount. A random scaling is applied to the covariance of each component; 
i.e. the eigenvectors and ratios between the eigenvalues of the covariance are the same as the ones of the Fisher matrix.

We obtain good results for shifts of about 0.5\% to 2\% of the box
size. Here, a trade-off between too large shifts (resulting in low
importance weights) and too small shifts (components stay near the
maximum, the posterior tails do not get sampled) has to be found.
The stretch factor is chosen randomly between typical values of
1 and 2. In some cases, in particular with high dimensionality, the
derivation of the Fisher matrix is not stable and the matrix is
numerically singular. In such cases we set the off-diagonal elements
of $F$ to zero.

We found a sample size between 7\,500 and 10\,000 points to be adequate
for most cases. The number of components $D$ of the initial importance function
was chosen between 5 to 10. For the final iteration we used a sample
size five times that of the initial sample size.

%%%%%%%%%%%%%%%%%%%%%%%%%%%%%%%%%%%%%%%%%%%%%%%%%%%%%%%%%%%%
\subsection{Results}
%%%%%%%%%%%%%%%%%%%%%%%%%%%%%%%%%%%%%%%%%%%%%%%%%%%%%%%%%%%%

%%%%%%%%%%%%%%%%%%%%%%%%%%%%%%%%%%%%%%%%%%%%%%%%%%%%%%%%%%%%
\subsubsection{General performance}
%%%%%%%%%%%%%%%%%%%%%%%%%%%%%%%%%%%%%%%%%%%%%%%%%%%%%%%%%%%%

\begin{figure}[!ht]

  \begin{center}
    \resizebox{\hsize}{!}{
      \includegraphics[bb=60 100 395 262]{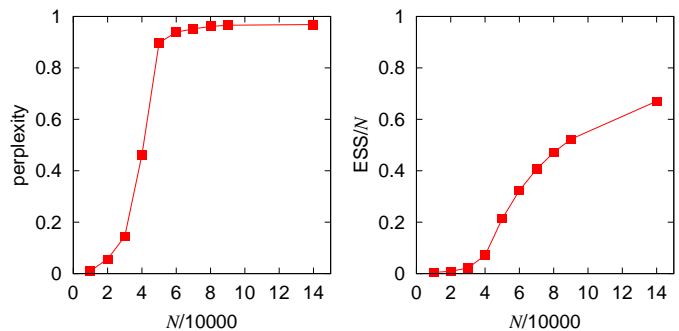}
    }
    \end{center}

    \caption{Perplexity (left panel) and normalised effective sample
      size ESS/$N$ (right panel), as a function of the cumulative
      sample size $N$. The likelihood is WMAP5 for a flat $\Lambda$CDM
      model with six parameters.  }
  \label{fig:perpl_cosmo_all}
\end{figure}

The PMC algorithm is reliable and very efficient in sampling and
exploring the parameter space. Both the perplexity as well as the
effective sample size increase quickly with each iteration
(Fig.~\ref{fig:perpl_cosmo_all}). The perplexity reaches values of
0.95 or more in many cases, although in particular for higher
dimensional posteriors the final values are lower. Satisfactory
results (i.e.~yielding consistent mean and marginals compared to MCMC,
see below) are obtained for perplexities larger than about 0.6.

\begin{figure}[!tb]

  \resizebox{\hsize}{!}{
    \includegraphics{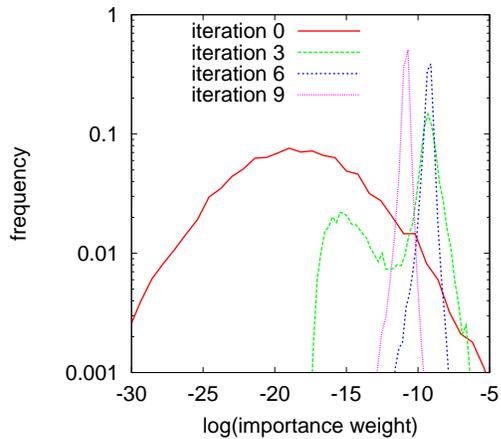}
  }

  \caption{Histogram of the normalised importance weights $\bar w_n^t$ for four
    iterations $t=0,3,6,9$. The posterior is WMAP5, flat $\Lambda$CDM model with
    six parameters.}
  \label{fig:weight_hist}
\end{figure}

The distribution of importance weights gets narrower from iteration to
iteration (Fig.~\ref{fig:weight_hist}). Initially, many sampled
points exhibit very low weights. After a few iterations, the
importance function has moved towards the posterior increasing the
efficiency of the sampling.

\begin{figure}[!tb]

  \begin{center}
    \resizebox{1.0\hsize}{!}{
      \includegraphics{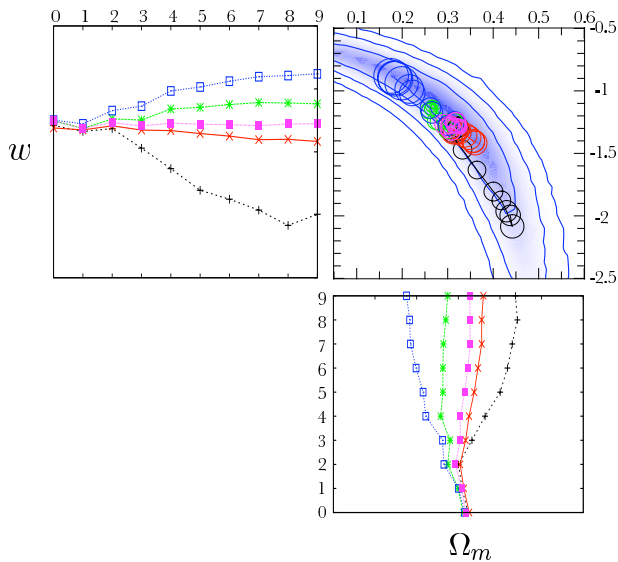}
    }
  \end{center}

  \caption{\emph{Lower left panel:} Overlaid to the SNIa confidence
    contours (68\%, 95\%, 99.7\%) is the movement of the importance
    function. For each iteration a circle is plotted at the position
    of the mean of each component, where different colours indicate
    different components.
    The circle size indicates the component weight.
    The starting point (first
    iteration, at $(0.3, -1.3)$) is marked by a thick circle.
    The other two panels show the mean positions in projection, fanned
    out as a function of the iteration.}
  \label{fig:SN-Omf-w+prop_mean}
\end{figure}

% La randonnee des composantes
Our initial mixture model starts with all mixture components close to
the maximum likelihood point. With consecutive iterations
the components spread out to better cover the region where the
posterior is significant. This can be seen in Fig.~\ref{fig:SN-Omf-w+prop_mean}.

\begin{figure}[!tb]

  \begin{center}
    \resizebox{0.8\hsize}{!}{
      \includegraphics{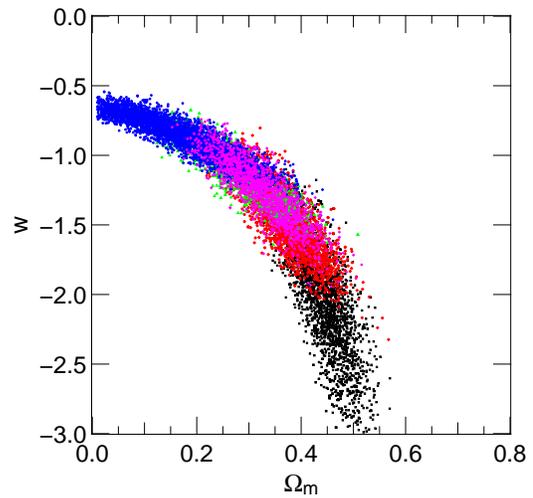}
    }
  \end{center}

  \caption{The sampled points from the final iteration are plotted,
    the colours indicate the components of the importance function
    from which the points are drawn (the colours are the same as in
    Fig.~\ref{fig:SN-Omf-w+prop_mean}). One out of five point is
    shown. Note that the density of points does not correspond to the
    posterior density since the former has to be weighted by the
    importance weights.}

  \label{fig_SN_proposal_means}

\end{figure}

Compared to traditional MCMC, our new PMC method is faster by orders
of magnitude. The time-consuming calculation of the posterior can be
performed in parallel and therefore a speed-up by a factor of the
number of CPUs is obtained. In times where clusters of
multi-core processors are readily available, this speed-up is easily
of the order of 100. In addition, MCMC has a low efficiency with
typical acceptance rates of 0.25 -- 0.3. The PMC normalised effective sample
size in the WMAP5 case is 0.7 which results in a much larger final
sample for the same number of posterior calculations of around
$150\,000$.

We emphasise again that with MCMC one can make
only limited use of parallel computing since one has to wait for each
Markov chain to converge, and because it is not straightforward to
combine chains, as mentioned earlier.

%%%%%%%%%%%%%%%%%%%%%%%%%%%%%%%%%%%%%%%%%%%%%%%%%%%%%%%%%%%%
\subsubsection{Comparison with MCMC}
\label{sec:comp_MCMC}
%%%%%%%%%%%%%%%%%%%%%%%%%%%%%%%%%%%%%%%%%%%%%%%%%%%%%%%%%%%%

The MCMC results we present here are either obtained using the adaptive MCMC
algorithm or a classical one. Indeed, as we show in the following, adaptive MCMC
can have some issues that a less efficient classical MCMC algorithm can avoid.
Apart from those special cases, the MCMC and adaptive MCMC gave very similar results,
the latter usually reaching a better acceptance rate, and thus a better efficiency.

We find excellent agreement between using our respective
implementations of MCMC (adaptive or not) and PMC. 
Mean, confidence intervals and
2d-marginals are very similar using both methods. The performance of
PMC is superior to MCMC in some cases, which is illustrated by the
following examples.

\begin{figure}[!tb]

  \resizebox{\hsize}{!}{
    \includegraphics{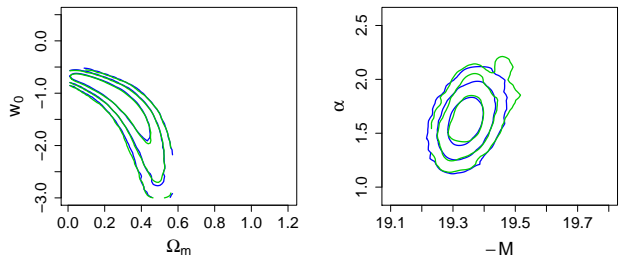}
  }

  \caption{Examples of marginalised likelihoods (68\%, 95\% and 99.7\%
    contours are shown) for PMC (solid blue) and MCMC (dashed green)
    from the SNIa data.}
  \label{fig:sn1a_2d}
\end{figure}

An inherent problem of MCMC is that even for a long run there
can be regions in parameter space that are not sampled in an unbiased
way. This is illustrated in Fig.~\ref{fig:sn1a_2d}. The feature at the
99.7\%-level of MCMC (left panel, for large values of $-M$ and
$\alpha$) is due to an ``excursion'' of the chain into a
low-likelihood region at step 130\,500, lasting for 300 steps. We ran
the chain for 300\,000 steps and the feature was still visible. A
second run of the chain did not exhibit this anomaly. This kind of sample
``noise'' can be prevented by running a chain for a very long
time or by combining several (converged) chains. Such features are
much less likely to occur in an importance sample which consists of
uncorrelated points.

\begin{figure}[!tb]

  \resizebox{0.8\hsize}{!}{
    % x-, y-labels changed in ps file: Helv <size> <size>,
    % size 600 -> 450
    \includegraphics{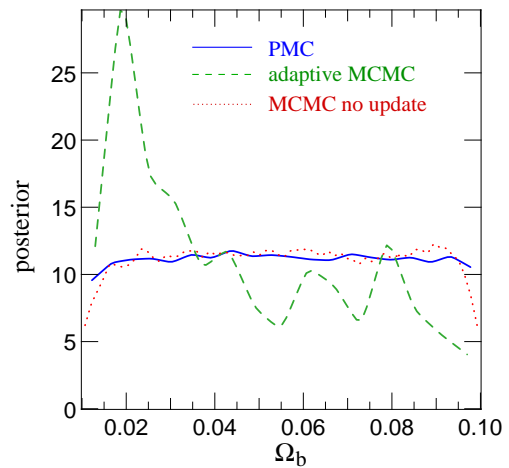}
  }

  \caption{Normalised 1d marginals for $\Omegab$ from weak lensing
    alone for PMC (solid blue lines) and MCMC (dashed green: adaptive,
    dotted red: non-adaptive).}
  \label{fig:L_Omegab}
\end{figure}

A second issue are parameters which are nearly unconstrained by the
data with the result that the marginalised posterior in that dimension
is flat. To illustrate this we choose weak lensing alone which can not
constrain $\Omegab$ (Fig.~\ref{fig:L_Omegab}). Using the Fisher matrix
as initial Gaussian proposal for adaptive MCMC, the chain stays
in a small region in the $\Omegab$-direction; the covariance being very flat, 
most jumps ends up out of the prior distribution.
This results in an update variance for this parameter which is much too small, 
and in a bad exploration of the posterior in this flat direction as shown 
Fig.~\ref{fig:L_Omegab}.
The classical MCMC algorithm, with the same proposal yields better results, 
but with a very low acceptance rate and needing 500\,000 steps to reach the 
result presented Fig.~\ref{fig:L_Omegab}. 
Alternatively, modifying the initial proposal to be smaller and better adapted 
to the prior, or increasing the covariance stretch factor from the optimal value 
of $c = 2.38^2/p$ (see Sect.~\ref{sec:propMCMC}) to $c = 3.2^2/p$ helps the chain to
explore more of the parameter space in the latter steps of the adaptation.
These modifications to the algorithm also result in very low acceptance rate, and somehow 
go against the very idea of an adaptive algorithm, since they require very 
fine tuning of the initial proposal!

With PMC we obtain a much better performance and recover very well the
flat posterior.

In Tables \ref{tab:mean_5} and \ref{tab:mean_5LS} we show the mean
and 68\% confidence intervals for CMB alone for the $\Lambda$CDM
model and for lensing+SNIa+CMB using $w$CDM, respectively. 
The differences in mean and 68\%-confidence intervals is less than a
few percent in most cases. Figure~\ref{fig:sn1a_2d} shows that the
lower-confidence regions and the correlation between parameters agrees
very well between (non-adaptive) MCMC and PMC.

\begin{table}
  \caption{Parameter means and 68\% confidence intervals for PMC and
    (non-adaptive) MCMC from the WMAP5 data.
  }
  \label{tab:mean_5}
  \renewcommand{\arraystretch}{1.5}
  \input{inputs/mean}
  \renewcommand{\arraystretch}{1}
\end{table}

\begin{table}
  \caption{Parameter means and 68\% confidence intervals for PMC using
    lensing, SNIa and CMB in combination. The (non-adaptive) MCMC
    results correspond to the values given in Table 5 from
    \cite{KB08}.}
  \label{tab:mean_5LS}
  \renewcommand{\arraystretch}{1.5}
  \input{inputs/mean_5LS}
  \renewcommand{\arraystretch}{1}
\end{table}

%%%%%%%%%%%%%%%%%%%%%%%%%%%%%%%%%%%%%%%%%%%%%%%%%%%%%%%%%%%%
\section{Discussion}
\label{sec:discussion}
%%%%%%%%%%%%%%%%%%%%%%%%%%%%%%%%%%%%%%%%%%%%%%%%%%%%%%%%%%%%

In this paper, we have introduced and assessed an adaptive importance
sampling approach, called Population Monte Carlo (PMC), which aims to
overcome the main difficulty in using importance sampling, namely the
reliance on a single efficient importance function. PMC achieves this
goal by iteratively adapting the importance function towards the
target density of interest.  A significant appeal of the approach,
when compared to alternatives such as MCMC, lies in the possibility to
use (massive) parallel sampling which considerably reduces the
computational time involved in the estimation of parameters for many
astrophysical and cosmological problems. Simulated and actual data
have been used in this work to assess the performance of PMC for
estimation of parameters in a Bayesian inference with features
approaching classical cosmological parameter posteriors.

The PMC approach is, in essence, an iterated importance sampling
scheme that simultaneously produces, at each iteration, a sample
approximately simulated from a target distribution $\pi$ and
an approximation of $\pi$ in the form of the current proposal distribution.
As such, the samples produced by the PMC approach can be exploited as
regular importance sampling outputs at any iteration $t$.  Samples
from previous iterations can be combined~\citep{AMIS:09}, and approximations
like $\hat \pi(f)$ can be updated dynamically, without necessarily
requiring the storage of samples.

Although adaptation of the importance function has the explicit aim of
improving the coverage of the posterior density there are instances
where this objective may not be met. In some cases, successive updates
of the importance function may result in: (a) an importance function
which is too peaked and which has light tails (invalid importance
function); (b) an importance function which fits only one mode (in the
case of a multimodal posterior); (c) numerical problems due to the
adaptation procedure (usually involving poor conditioning of some of
the covariance matrices). Such cases are likely to produce a poor
approximation to the integral of interest, or alternatively lead to
highly variable parameter estimates over iterations. These problems
can be quickly discovered or signalled by observing a poor ESS, and
parameter estimates or normalised perplexity which do not stabilise
after a few iterations.

Such cases of poor performance as outlined can be significantly reduced by
choosing a reasonably well informed initial importance function with
a large enough sample size at each iteration, especially on the
initial iteration that requires many points to counter-weight a
potentially poor importance function. In general, the initial
importance function should be chosen to cover a region of the
parameter space that has support larger than the posterior. In the
absence of reliable prior information, finding such an importance
function may be difficult to do.  One approach may be to locate the
components in the centre of the feasible range (if available) for each
variable, with reasonably large variances to ensure some coverage of
the parameter space.  We found this approach to be reasonably
successful for the simulated data case discussed in Sect.~\ref{sec:simul}.  In the
presence of some prior information, for example an estimate of the maximum-likelihood
point and an approximation of the covariance matrix (via the Hessian),
components can be placed around these points with variance
comparable to the approximation. Another approach may be to
perform a singular value decomposition of the covariance matrix, and
make use of the eigenvectors and eigenvalues to place components along
the most likely directions of interest.  Alternatively and in the same
spirit, components can be placed according to the principal points of
the resulting sample, using a k-means clustering approach
\citep{tarpey07}. Both approaches have been reasonably successful for
a range of posterior densities examined, and by placing components in
regions of high posterior support in addition to the mode have the
potential to further significantly reduce the number of iterations for
difficult posterior densities.

The main appeals or advantages of the PMC method are worth
re-emphasising at this point:

\begin{enumerate}

  \item Parallelisation of the posterior calculations

  \item Low variance of Monte Carlo estimates

  \item Simple diagnostics of `convergence' (perplexity)

\end{enumerate}

We address these three points in more detail now.

(1.) The first advantage, namely the ability to parallelise the computational
task, is becoming increasingly useful through the availability of
cheap multi-CPU computers and the standardization of clusters of
computers. Software to implement the parallelisation task, such as
Message Parsing Interface (MPI)
\footnote{http://www-unix.mcs.anl.gov/mpi/}, are publicly available
and relatively straightforward to implement. For the cosmological
examples presented (Sect.~\ref{sec:cosmo}), we used up to 100 CPUs on
a computer cluster to explore the cosmology posteriors. In the case of
WMAP5, this reduced the computational time from several days for MCMC
to a few hours using PMC.

(2.) In general, for PMC and an importance function that is closely
matched to the target density, significant reductions in the variance
of the Monte Carlo estimates are possible in comparison to estimates
obtained using MCMC \citep{robert:casella:2004}.  For example, for the
posterior estimates for the WMAP5 data we observed a 10-fold reduction in variance for the same number of sample points as observed for MCMC. Such reductions suggest that the computational time savings extend not only to the number of CPU's available but to smaller sample requirements for PMC in total  compared to MCMC to achieve similar variability of estimates. For cosmological applications, this observation is valuable as we observed, e.g., in Fig.~\ref{fig:perpl_cosmo_all} for CMB data, that the fit between the adapted importance function and the target posterior is sometimes quite good.  By combining samples across iterations further computational savings are also possible. The absence of construction of a Markov chain for PMC can also have the desirable attribute of reducing sample noise, as observed for the SNIa data in Sect.~\ref{sec:comp_MCMC}.

(3.) As shown in Sect.~\ref{sec:PMC_monitor}, the perplexity
(eq.~\ref{perplexity}) is a relatively simple measure of sampling
adequacy to the target density of interest.  For MCMC and other
approaches which rely on formal measures of convergence, assessment of
convergence can be very difficult with users facing a potential array
of associated diagnostic tests.

In addition to the above points, a further appeal of PMC is the ability to provide a very good
approximation to the marginal posterior or evidence, which
naturally follows as a byproduct of the approach. 
To demonstrate this
appeal, further research is underway to explore the use of PMC in the
context of model selection problems in cosmology.

%%%%%%%%%%%%%%%%%%%%%%%%%%%%%%%%%%%%%%%%%%%%%%%%%%%%%%%%%%%%
\begin{acknowledgments}
%%%%%%%%%%%%%%%%%%%%%%%%%%%%%%%%%%%%%%%%%%%%%%%%%%%%%%%%%%%%

We acknowledge the use of the Legacy Archive for Microwave Background
Data Analysis (LAMBDA). Support for LAMBDA is provided by the NASA
Office of Space Science. We thank the Planck group at IAP and the
\textsc{Terapix} group for support and computational facilities. DW and
MK are supported by the CNRS ANR ``ECOSSTAT'', contract number
ANR-05-BLAN-0283-04 ANR ECOSSTAT. The authors would like to thank
F.~Bouchet, S.~Bridle, J.-M.~Marin, Y.~Mellier and I.~Tereno for helpful discussions.

%%%%%%%%%%%%%%%%%%%%%%%%%%%%%%%%%%%%%%%%%%%%%%%%%%%%%%%%%%%%
\end{acknowledgments}
%%%%%%%%%%%%%%%%%%%%%%%%%%%%%%%%%%%%%%%%%%%%%%%%%%%%%%%%%%%%

%%%%%%%%%%%%%%%%%%%%%%%%%%%%%%%%%%%%%%%%%%%%%%%%%%%%%%%%%%%%
\begin{appendix}
%%%%%%%%%%%%%%%%%%%%%%%%%%%%%%%%%%%%%%%%%%%%%%%%%%%%%%%%%%%%

%%%%%%%%%%%%%%%%%%%%%%%%%%%%%%%%%%%%%%%%%%%%%%%%%%%%%%%%%%%%
\section{Details of the importance function updates for PMC}
\label{sec:app-update-PMC}
%%%%%%%%%%%%%%%%%%%%%%%%%%%%%%%%%%%%%%%%%%%%%%%%%%%%%%%%%%%%

The method proposed in \cite{cappe:douc:guillin:marin:robert:2007} to adaptively update the
parameters of the importance function $q^t$ is based on a variant of the EM (Expectation-Maximization) algorithm
\cite{dempster:laird:rubin:1977}, which is the standard tool for the estimation of the parameters
of mixture densities. We describe below the principle underlying the algorithm of
\cite{cappe:douc:guillin:marin:robert:2007}, showing in particular that each iteration
decreases, up to the importance sampling approximation errors, the Kullback divergence between the target $\pi$ and the importance function $q^t$.

Remember that our goal is to minimise~\eqref{eqn:kdiv}, as $t$ increases, by iteratively tuning the parameters $\alpha^t,\theta^t$ of the mixture importance function defined in~\eqref{eq:mixtureISdensity}. Developing the logarithm in (\ref{eqn:kdiv}), this objective can be equivalently formulated in terms of the maximization of the following quantity
\begin{equation}
 \ell(\alpha,\theta) = \int \text{log}\left(\sum_{d=1}^D \alpha_d \varphi(x; \theta_d)\right) \pi(x) \,\dd x
\label{eqn:obsdiv}
\end{equation}
with respect to $\alpha,\theta$. Using Bayes' rule, we denote by
\begin{equation}
\rho_{d}(x;\alpha,\theta) = \frac{\alpha_d \varphi(x;\theta_d)}{\sum_{d=1}^{D}\alpha_d \varphi(x;\theta_d)}
\label{eq:postprobmix}
\end{equation}
the posterior probability that $x$ belongs to the $d$th component of the mixture (for the mixture parameters $\alpha,\theta$). The EM principle consists of evaluating, at iteration $t$, the
following intermediate quantity
\begin{equation}
 L^{t}(\alpha,\theta) = \int \sum_{d=1}^D \rho_{d}(x;\alpha^{t},\theta^{t})\text{log}\left(\alpha_d \varphi(x; \theta_d)\right) \pi(x) \,\dd x.
\label{eqn:intquant}
\end{equation}
Using the concavity of the log as well as the expression of $\rho_{d}$ in \eqref{eq:postprobmix}, it is easily checked that
\begin{multline}
  \sum_{d=1}^D \rho_{d}(x;\alpha^{t},\theta^{t})\text{log}\left(\frac{\alpha_d \varphi(x; \theta_d)}{\alpha_d^{t} \varphi(x; \theta_d^{t})}\right) \\
  \leq \log\left(\frac{\sum_{d=1}^D \alpha_d \varphi(x; \theta_d)}{\sum_{d'=1}^D \alpha_{d'}^{t} \varphi(x; \theta_{d'}^{t})}\right)
  \label{eq:random:lfhce}
\end{multline}
and hence that $L^{t}(\alpha,\theta) - L^{t}(\alpha^{t},\theta^{t}) \leq \ell(\alpha,\theta) -
\ell(\alpha^{t},\theta^{t})$. Thus, any value of $\alpha,\theta$ which increases the intermediate
quantity $L^{t}$ above the level $L^{t}(\alpha^{t},\theta^{t})$ also results in, at least,
an equivalent increase of the actual objective function $\ell$. In the EM algorithm, one sets
$\alpha^{t+1}$ and $\theta^{t+1}$ to the values where the intermediate quantity $L^{t}(\alpha,\theta)$ is
maximal, thus satisfying the previous requirement. Furthermore, the maximization of
$L^{t}(\alpha,\theta)$ leads to a closed form solution whenever $\varphi$ belongs to a
so-called exponential family of probability densities.

In the example of the multivariate Gaussian density recalled in \eqref{eq:gaussian}, the parameter $\theta_d$ consists of the mean $\mu_d$ and the covariance matrix $\Sigma_d$ and the
intermediate quantity may be written as
\begin{multline}
 L^{t}(\alpha,\mu,\Sigma) = \int \sum_{d=1}^D \rho_{d}(x;\alpha^{t},\mu^{t},\Sigma^{t}) \bigg\{ \log(\alpha_d) \\
   -\frac{1}{2}\left[\log|\Sigma_d| + (x-\mu_{d})^\transp \Sigma_{d}^{-1}(x-\mu_{d})\right]\bigg\}  \pi(x) \,\dd x ,
\label{eqn:intquant:gauss}
\end{multline}
up to terms that do not depend on $\alpha$, $\mu$ or $\Sigma$.
Routine calculations show that the maximum of~\eqref{eqn:intquant:gauss} is achieved for
\begin{align}
  & \alpha_d^{t+1}=\int \rho_d(x;\alpha^{t},\mu^{t},\Sigma^t) \pi(x) \dd x, \\
  & \mu_d^{t+1} = \frac{\int x \rho_d(x;\alpha^{t},\mu^{t},\Sigma^t) \pi(x) \dd x}{\alpha_d^{t+1}}, \\
  & \Sigma_d^{t+1} = \frac{\int (x-\mu_d^{t+1})(x-\mu_d^{t+1})^\transp \rho_d(x;\alpha^{t},\mu^{t},\Sigma^t) \pi(x) \dd x}{\alpha_d^{t+1}}.
\end{align}
In practice, both the numerator and denominator of each of the above expressions are integrals under
$\pi$ which must be approximated. The solution proposed in
\cite{cappe:douc:guillin:marin:robert:2007} is based on self-normalised importance sampling using
the weighted sample simulated at the previous iteration $(x_1^{t}, \bar{w}_1^{t}), \dots, (x_N^{t},
\bar{w}_N^{t})$. The corresponding empirical update equations are
given in eqs.~\eqref{eq:update_alpha}--\eqref{eq:update_Sigma} of
Sect.~\ref{eq:gaussian_update}.

The Student-t distribution provides a family of multivariate densities
with parameters $\mu$ and $\Sigma$ which have the same interpretation
as in the Gaussian case (except for the fact that the covariance is
equal to $\nu/(\nu-2)\Sigma$ rater than $\Sigma$) but with an
additional shape factor $\nu \geq 2$ which allow for heavier tails:
letting $\nu \to \infty$ yields back the Gaussian but for $\nu = 2$,
one obtains a density with polynomially decreasing tails whose only
finite moments are the two first ones (note that it is also possible to extend the
family to the case where $0 < \nu < 2$). Using mixtures of Student-t
distributions will thus be mostly useful in cases where the target
posterior distribution $\pi$ itself has heavy tails. The parameter
update corresponding to mixtures of Student-t distributions is a bit
more involved but follows the same general pattern.  For the sake of
completeness, we just recall below the formulas given
in~\cite{cappe:douc:guillin:marin:robert:2007}:
\begin{align*}
 & \alpha^{t+1}_{d} = \sum_{n=1}^N\bar{w}_n^t \rho_{d}(x_n^t;\alpha^{t},\theta^{t}), \\
 & \mu^{t+1}_{d} = \frac{\sum_{n=1}^N\bar{w}_n^t \rho_{d}(x_n^t;\alpha^{t},\theta^{t})\gamma_{d}(x_n^t;\theta^{t})x_n^t}{\sum_{n=1}^N\bar{w}_n^t\rho_{d}(x_n^t;\alpha^{t},\theta^{t})\gamma_{d}(x_n^t;\theta^{t})} \\
 & \Sigma^{t+1}_{d} = \frac{1}{\sum_{n=1}^N\bar{w}_n^t\rho_{d}(x_n^t;\alpha^{t},\theta^{t})} \times \\
 & \, \sum_{n=1}^N\bar{w}_n^t\rho_{d}(x_n^t;\alpha^{t},\theta^{t})\gamma_{d}(x_n^t;\theta^{t})(x_n^t-\mu^{t+1}_{d})(x_n^t-\mu^{t+1}_{d})^\transp
\end{align*}
where
\begin{equation}
\rho_{d}(x;\alpha,\theta) = \frac{\alpha_d \tau(x;\nu_{d},\mu_{d},\Sigma_{d})}{\sum_{d=1}^{D}\alpha_d \tau(x;\nu_{d},\mu_{d},\Sigma_{d})};
\end{equation}
\begin{equation}
\gamma_{d}(x_n^t;\theta) = \frac{\nu_{d}+p}{\nu_{d}+(x-\mu_{d})^\transp(\Sigma_{d})^{-1}(x-\mu_{d})}
\end{equation}
and $\tau(\cdot;\mu,\Sigma,\nu)$ denotes the $p$-dimensional Student-t probability density function,\begin{multline}
  \tau(x;\mu,\Sigma,\nu) = \frac{\Gamma((\nu+p)/2)}{\Gamma(\nu/2)\nu^{p/2}\pi^{p/2}} |\Sigma|^{-1/2}\times \\
  \left(1+\frac{1}{\nu} (x-\mu)^\transp\Sigma^{-1}(x-\mu)\right)^{-(\nu+p)/2}.
\end{multline}

Sampling from a multivariate Student-t distribution is most easily
undertaken by using its derivation in terms of a multivariate
Gaussian ($Y\sim N_k(0,\Sigma)$) and chi-squared distribution
($Z\sim\chi^{2}_\nu$),
\begin{equation*}
x=\mu+y\sqrt{\nu/z}
\end{equation*}
and taking advantage of the fact that sampling from Y and Z is
straightforward.

%%%%%%%%%%%%%%%%%%%%%%%%%%%%%%%%%%%%%%%%%%%%%%%%%%%%%%%%%%%%
\end{appendix}
%%%%%%%%%%%%%%%%%%%%%%%%%%%%%%%%%%%%%%%%%%%%%%%%%%%%%%%%%%%%

% version bibtex
%\bibliographystyle{apsrev}
%\bibliography{big}
% Produces the bibliography via BibTeX.

%version arxiv

\input{inputs/paper1.bbl}
\newpage

\end{document}

%% file: inputs/abbr-journals.tex
\def\apj{ApJ}%
\def\apjl{ApJ}%
\def\apjs{ApJS}%
\def\aap{A\&A}%
\def\mnras{MNRAS}%

%% file: inputs/table_simresults.tex
\begin{tabular}{cccc}
    \hline
      & & PMC & MCMC \tabularnewline
    \hline
    $\pi(f_{a})$ & mean & 0.097 & -0.028 \tabularnewline
    " & std & 0.218  & 0.536 \tabularnewline
    $\pi(f_{b})$ & mean & 0.013 & 0.002 \tabularnewline
    " & std & 0.163 & 0.315 \tabularnewline
    \hline
    acceptance & & - &  0.11 \tabularnewline
    perplexity & & 0.80 & - \tabularnewline
    \hline
\end{tabular}

%% file: inputs/mean.tex
\begin{center}\begin{tabular}{|l|l|l|}\hline
\rule[-3mm]{0em}{8mm}Parameter	 & PMC	 & MCMC	\\ \hline\hline
$\Omega_{\rm\,b}$	 & $0.04424^{+0.00321}_{-0.00290}$	 & $0.04418^{+0.00321}_{-0.00294}$	\\ \hline
$\Omega_{\rm\,m}$	 & $0.2633^{+0.0340}_{-0.0282}$	 & $0.2626^{+0.0359}_{-0.0280}$	\\ \hline
$\tau$	 & $0.0878^{+0.0181}_{-0.0160}$	 & $0.0885^{+0.0181}_{-0.0160}$	\\ \hline
$n_{\rm\,s}$	 & $0.9622^{+0.0145}_{-0.0143}$	 & $0.9628^{+0.0139}_{-0.0145}$	\\ \hline
$10^9\Delta^2_R$	 & $2.431^{+0.118}_{-0.113}$	 & $2.429^{+0.123}_{-0.108}$	\\ \hline
$h$	 & $0.7116^{+0.0271}_{-0.0261}$	 & $0.7125^{+0.0274}_{-0.0268}$	\\ \hline
\end{tabular}\end{center}

%% file: inputs/mean_5LS.tex
\begin{center}\begin{tabular}{|l|l|l|}\hline
\rule[-3mm]{0em}{8mm}Parameter	 & PMC	 & MCMC	\\ \hline\hline
$\Omega_{\textrm{b}}$	 & $0.0432^{+0.0027}_{-0.0024}$	 & $0.0432^{+0.0026}_{-0.0023}$	\\ \hline
$\Omega_{\textrm{m}}$	 & $0.254^{+0.018}_{-0.017}$	 & $0.253^{+0.018}_{-0.016}$	\\ \hline
$\tau$	 & $0.088^{+0.018}_{-0.016}$	 & $0.088^{+0.019}_{-0.015}$	\\ \hline
$w$	 & $-1.011\pm{0.060}$	 & $-1.010^{+0.059}_{-0.060}$	\\ \hline
$n_{\textrm{s}}$	 & $0.963^{+0.015}_{-0.014}$	 & $0.963^{+0.015}_{-0.014}$	\\ \hline
$10^9\Delta^2_R$	 & $2.413^{+0.098}_{-0.093}$	 & $2.414^{+0.098}_{-0.092}$	\\ \hline
$h$	 & $0.720^{+0.022}_{-0.021}$	 & $0.720^{+0.023}_{-0.021}$	\\ \hline
$a$	 & $0.648^{+0.040}_{-0.041}$	 & $0.649^{+0.043}_{-0.042}$	\\ \hline
$b$	 & $9.3^{+1.4}_{-0.9}$	 & $9.3^{+1.7}_{-0.9}$	\\ \hline
$c$	 & $0.639^{+0.084}_{-0.070}$	 & $0.639^{+0.082}_{-0.070}$	\\ \hline
$-M$	 & $19.331\pm{0.030}$	 & $19.332^{+0.029}_{-0.031}$	\\ \hline
$\alpha$	 & $1.61^{+0.15}_{-0.14}$	 & $1.62^{+0.16}_{-0.14}$	\\ \hline
$-\beta$	 & $-1.82^{+0.17}_{-0.16}$	 & $-1.82\pm{0.16}$	\\ \hline
$\sigma_8$	 & $0.795^{+0.028}_{-0.030}$	 & $0.795^{+0.030}_{-0.027}$	\\ \hline
\end{tabular}\end{center}